\documentclass[longbibliography,twocolumn]{revtex4-2}

\usepackage{amsmath}
\usepackage{graphicx}

\def\nn{\nonumber}

\begin{document}

\title{Introducing Corrections to the Reflectance of Graphene by Light Emission}

\author{Ken-ichi Sasaki}
\email{Corresponding author: ke.sasaki@ntt.com}
\affiliation{NTT Research Center for Theoretical Quantum Physics and NTT Basic Research Laboratories, NTT Corporation,
3-1 Morinosato Wakamiya, Atsugi, Kanagawa 243-0198, Japan}

\author{Kenichi Hitachi}
\affiliation{NTT Basic Research Laboratories, NTT Corporation,
3-1 Morinosato Wakamiya, Atsugi, Kanagawa 243-0198, Japan}

\author{Masahiro Kamada}
\affiliation{Advanced Research Laboratory, Anritsu Corporation, 5-1-1 Onna, Atsugi, Kanagawa 243-8555, Japan}

\author{Takamoto Yokosawa}
\affiliation{Advanced Research Laboratory, Anritsu Corporation, 5-1-1 Onna, Atsugi, Kanagawa 243-8555, Japan}

\author{Taisuke Ochi}
\affiliation{Advanced Research Laboratory, Anritsu Corporation, 5-1-1 Onna, Atsugi, Kanagawa 243-8555, Japan}

\author{Tomohiro Matsui}
\affiliation{Advanced Research Laboratory, Anritsu Corporation, 5-1-1 Onna, Atsugi, Kanagawa 243-8555, Japan}

\date{\today}

\begin{abstract}
 Monolayer graphene absorbs 2.3 percent of the incident visible light. 
 This “small” absorption has been used to emphasize the visual transparency of graphene, 
 but it in fact means that multilayer graphene absorbs a sizable fraction of incident light, 
 which causes non-negligible fluorescence.
 In this paper, 
 we formulate the light emission properties of multilayer graphene composed of tens to hundreds of layers
 using a transfer matrix method and confirm the method's validity experimentally.
 We could quantitatively explain the measured contrasts of multilayer graphene on SiO$_2$/Si substrates
 and found sizable corrections, which cannot be classified as incoherent light emissions,
 to the reflectance of visible light.
 The new component originates from coherent emission caused by absorption at each graphene layer.
 Multilayer graphene thus functions as a partial coherent light source of various wavelengths,
 and it may have surface-emitting laser applications. 
\end{abstract}


\maketitle

\section{Introduction}

The visual detection of graphene on Si substrates involves a complex scientific aspect 
related to the interference effects of light.~\cite{Novoselov2005,zhang2005,Novoselov2005a}
To enhance visibility, it is crucial for the contrast 
between the reflectances from the substrate and the graphene layer to be sufficiently large.
Si substrates with a specific thickness of ${\rm SiO_2}$ ($d_{\rm SiO_2}$) 
are known to provide advantages in this context.~\cite{Blake2007a,Roddaro2007a}
Specifically, when the optical path length difference of the ${\rm SiO_2}$ layer is 
one-half or three-halves of the wavelength of the incident light, the contrast reaches its maximum.

The enhanced visibility of graphene is primarily attributed to a substrate-induced enhancement of light absorption.
This enhancement is valuable not only for graphene detection but also for exploring notable phenomena. 
For instance, the reflectance of specific graphene multilayers on SiO$_2$/Si substrates can be reduced to zero 
for normally incident visible light with a wavelength ($\lambda$) approximately equal to $2d_{\rm SiO_2}$. 
Zero reflectance is achieved through destructive interference caused by ${\rm SiO_2}$ 
and a significant absorption coefficient of graphene.~\cite{H.Ni2007,Casiraghi2007,Ghamsari2016}
More importantly, the reflectance of multilayer graphene on SiO$_2$/Si substrates is intricately determined 
because the significant absorption may result in non-negligible luminescence. 
This luminescence also functions as a secondary light source, 
leading to more sophisticated interference effects of light.~\cite{Nemanich1980,Wang2008,Stohr2010,Lui2010,Liu2010,Suemoto2013,Riaz2015}

Since graphene lacks an energy bandgap, unlike semiconductors, 
it is naturally expected that the impact of light emission on its optical properties is negligible. 
However, electrons that have absorbed light undergo 
various relaxation processes towards the ground state,~\cite{Song2013b,Massicotte2021}
generally involving luminescence contributions. 
For instance, Lui {\it et al.}~\cite{Lui2010} and Suemoto {\it et al.}~\cite{Suemoto2013}
have reported that graphene and graphite emit light under femtosecond laser pulse excitation, respectively.
In this paper, we demonstrate that, despite the tiny branching ratio of light emission to absorption, 
light emission constitutes the primary correction to visible reflectance, thanks to the substantial absorption of graphene.
We have found this notable feature for the first time
by leveraging both the destructive interference effects of substrates and the coherence of the multilayer structure.
Additionally, we can show that light emission significantly contributes to enhancing the visibility of graphene on a substrate.

In this study, 
we delineate coherent and incoherent corrections to 
the reflectance of multilayer graphene (on SiO$_2$/Si substrates)
caused by the light it emits after absorbing incident light. 
The coherent corrections are dependent on only two parameters:
the branching ratio (${\cal B}$) of coherent light emission to absorbed light
and the phase ($+$ or $-$) of the emitted light.
Both parameters are assumed to be independent of the number of layers ($N$)
and the wavelength of light.
Contrasts (reflectivities) calculated using these phenomenologically determined parameters
show reasonable agreement with measured values for various $N$.
Despite the branching ratio, determining the strength of light emitted from a graphene monolayer, 
being less than one percent, the coherent components become the primary corrections to the reflectance of multilayer graphene. 
This is due to the coherence increasing the amplitude of reflection through constructive interference. 
Furthermore, based on the observation that the phase of reflection and 
transmission coefficients of each layer translates into the amplitude of the emitted light, 
we hypothesize that the origin of the coherent components is coherent light emission stimulated by the incident light. 
Meanwhile, the Raman effect is explained as incoherent corrections.
These conclusions are independent of the specific values of the two parameters.

Recently, the optical constants of single-layer graphene were measured with high accuracy.~\cite{Castriota2019,El-Sayed2021}
However, classical electromagnetic models were employed for fitting and interpreting the results.
Our approach advances the physical understanding of the measured optical constants of 
single-layer graphene to a more fundamental level, 
incorporating principles of quantum electrodynamics that describe the creation of photons (light emission). 
Moreover, our formulation is versatile and generally applicable to any layered material and superlattice, 
promising an accurate description of their optical processes.

Multilayer graphene composed of tens to hundreds of layers
is an interesting research subject.
However, it has not been explored much, partly because 
the success of the exfoliation method has rapidly shifted the interest of many researchers 
from infinite layers of graphite to few-layer graphene.
The obvious advantage of multilayer graphene is that it can increase the signal strength,~\cite{Devang2023}
whereas the signal strength of few-layer graphene is low and difficult to measure.
Besides that, multilayer graphene hosts various intriguing phenomena.
For instance, studies have revealed that the optimal number of layers for absorbing infrared radiation is 87.~\cite{Sasaki2020a}
Additionally, a notable nonlinear optical effect has been attributed to multilayer graphene.~\cite{Hendry2010,Sun2010,Mikhailov2012}
Yang {\it et al.}, 
for instance, demonstrated the maximum third-harmonic signal from 24-layer graphene on a quartz substrate.~\cite{Yang2018}

This paper is structured as follows: 
In Sec.~\ref{sec:sec2}, 
we present fundamental insights into the reflectance of multilayer graphene on SiO$_2$/Si substrates. 
By utilizing the Fresnel equation, we can replicate measured contrasts and assert that 
the optical constants obtained by El-Sayed {\it et al.}~\cite{El-Sayed2021} 
sufficiently and accurately describe the reflectance. 
However, it is noteworthy that these optical constants 
lead to an unexpectedly large effective coupling between light and graphene, 
a phenomenon inconsistent with existing experiments.
Section~\ref{sec:sec3} introduces our formulation of corrections to the reflectance arising from light emissions.
Our model effectively describes the reflectance without introducing such inconsistencies. 
These corrections are categorized into two types: 
coherent emission, corresponding to stimulated emission with a common phase, 
and incoherent emission, featuring a random phase identified as the Raman effect. 
A detailed comparison between calculated and measured contrasts is presented in Sec.~\ref{sec:sec4}. 
Finally, Section~\ref{sec:dis} offers a discussion of the findings, 
and Section~\ref{sec:con} provides the conclusions.

\

\section{Reflectance of graphene multilayer on Si substrate}\label{sec:sec2}

In this section, we demonstrate
that the measured reflectance of various graphene multilayers with different thickness 
aligns well with the Fresnel equation when adopting
optical constants ($n$ and $k$) obtained by El-Sayed {\it et al.}~\cite{El-Sayed2021} 
through ellipsometric measurements of chemical vapor deposited (CVD) graphene monolayer.
The key observation from this section is that 
electron-light coupling constant, extracted from the established $n$ and $k$ values, 
deviates significantly from the fine-structure constant $\alpha\simeq 1/137$.
In fact, the difference is beyond the level of small corrections of order of $\alpha^2$ caused by 
such as a change in the band dispersion at high energy from linear dispersion 
(commonly known as trigonal warping effects) and Fermi velocity renormalization,~\cite{Stauber2008,Stauber2017} 
which have been discussed theoretically thus far.
This observation partly motivates the introduction of a new reflectance model developed in the subsequent sections.

\subsection{Fresnel equation}

The reflectance of $N$-layer graphene on SiO$_2$/Si substrate to normally incident light of wavelength $\lambda$ 
can be formulated using the reflection coefficient $r_N(\lambda)$ as $R_N(\lambda) = |r_N(\lambda)|^2$
(Fresnel equation),
where
\begin{widetext}
\begin{small}
\begin{align}
 r_N(\lambda) = 
 \frac{\left[  (1-n_{\rm Si})\cos\varphi - i \left( 
 \frac{n_{\rm Si}}{n_{\rm SiO_2}} -n_{\rm SiO_2} \right) \sin \varphi \right] \cos \phi 
 -\left[ \left( \frac{n_{\rm SiO_2}}{\sqrt{\varepsilon_{\rm g}}} - \frac{n_{\rm Si} \sqrt{\varepsilon_{\rm g}}}{n_{\rm SiO_2}}
 \right) \sin \varphi + i 
 \left( \frac{n_{\rm Si}}{\sqrt{\varepsilon_{\rm g}}} - \sqrt{\varepsilon_{\rm g}} \right) \cos \varphi 
 \right] \sin \phi}{ \left[
(1+n_{\rm Si})\cos\varphi - i \left(   \frac{n_{\rm Si}}{n_{\rm SiO_2}} + n_{\rm SiO_2} \right)\sin \varphi \right]\cos \phi
 -\left[ \left( \frac{n_{\rm SiO_2}}{\sqrt{\varepsilon_{\rm g}}} + \frac{n_{\rm Si}\sqrt{\varepsilon_{\rm g}}}{n_{\rm SiO_2}}
 \right) \sin \varphi + i 
 \left( \frac{n_{\rm Si}}{\sqrt{\varepsilon_{\rm g}}} + \sqrt{\varepsilon_{\rm g}} \right) \cos \varphi 
 \right] \sin \phi
 }.
\label{eq:RN}
\end{align}
\end{small}
\end{widetext}
Here,
$\varphi \equiv n_{\rm SiO_2} d_{\rm SiO_2} \frac{2\pi}{\lambda}$
is the phase acquired by light after it propagates through a distance $d_{\rm SiO_2}$
in SiO$_2$, and $\phi \equiv \sqrt{\varepsilon_{\rm g}} \frac{2\pi}{\lambda}d_N$
is the complex phase acquired when light passes through $N$-layer graphene of thickness $d_N \equiv Nd$.
Multilayer graphene is treated as an effective medium whose unit length is the interlayer spacing $d$ ($=0.335$ nm)
and its dielectric constant is given by $\varepsilon_g=(n+ik)^2$.~\cite{El-Sayed2021}
$n_{\rm Si}$ and $n_{\rm SiO_2}$ are the refractive indexes of Si and SiO$_2$, respectively.
Si is treated as an absorbing substrate having a semi-infinite thickness ($n_{\rm Si}$ is a complex number)
whose dispersion is taken into account,~\cite{Aspnes1983a}
while SiO$_2$ is treated as an absorption-free film.~\cite{Malitson1965}
$R_N(\lambda)$ depends sensitively on the two phases, $\varphi$ and $\phi$.
When $N=0$ or $\phi=0$ in Eq.~(\ref{eq:RN}), 
$R_0(\lambda)$ corresponds to the reflectance of the substrate without graphene.
It can be minimized for a specific $\lambda$ by destructive interference; namely, 
$R_0(\lambda)$ is at a minimum when $\cos\varphi=0$
as $\left|\frac{n_{\rm Si}-n^2_{\rm SiO_2}}{n_{\rm Si}+n^2_{\rm SiO_2}} \right|^2$.~\cite{Anders1965}
Monolayer graphene is most easily detectable on SiO$_2$/Si substrates 
when destructive interference occurs, because 
$|R_0(\lambda)-R_1(\lambda)|$ takes a maximum when $\cos\varphi=0$
(i.e. when $\lambda\simeq 2d_{\rm Si0_2}$ because $n_{\rm SiO_2}\simeq 1.46$).~\cite{Blake2007a}

\subsection{Comparison of measured and calculated contrasts}

Multilayer graphene was prepared by exfoliating highly oriented pyrolytic graphite
(HOPG) on the same SiO$_2$/Si substrate.
The reflectance of the multilayer graphene was measured 
with a spectroscopic reflectometer (TohoSpec3100, Toho Technology) using a $\times 50$ objective lens.
First, we determined that $d_{\rm SiO_2}=268$ nm
from the reflectance of the substrate (Appendix~\ref{app:substrate}).
This value is used consistently in all the calculations reported in this paper,
and it results in that destructive interference occurs for $R_0$ at $\lambda\simeq 520$ nm.
The thickness of the graphene flakes was determined by atomic force microscopy (Dimension XR, Bruker).

\begin{figure}[htp]
 \begin{center}
  \includegraphics[scale=0.6]{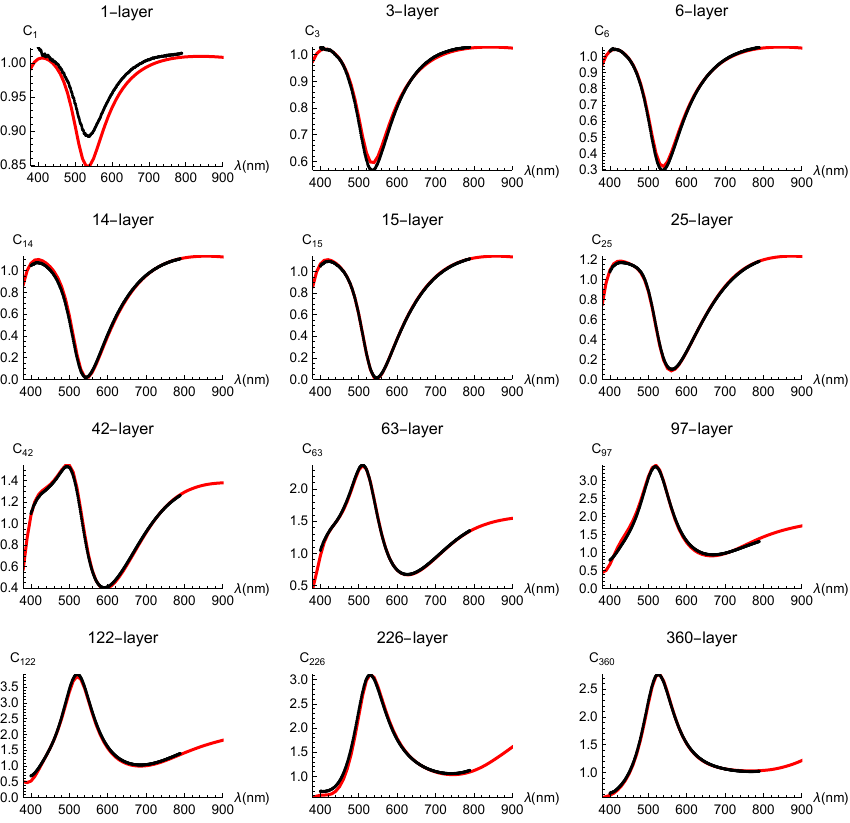}
 \end{center}
 \caption{{\bf Measured (black circles) and calculated (red solid) spectral contrasts
 of $N$-layer graphene on the same SiO$_2$/Si substrate.}
 The measurments were conducted using a white-light source at room temperature. 
 The red solid curves were obtained from Eq.~(\ref{eq:RN}) with $\varepsilon_g = (n+ik)^2$
 using optical constants obtained by El-Sayed {\it et al.}~\cite{El-Sayed2021}
 The horizontal axis is $\lambda$ (nm), and the reliable range of our spectrometer is 450 to 800 nm.
 }
 \label{fig:measured_fresnel}
\end{figure}

The representative measured spectral contrasts
are depicted as black dots (circles) in Fig.~\ref{fig:measured_fresnel}
(The error bars for the data are within each circle).
It is important to note that 
we present contrasts ($C_N\equiv R_N/R_0$) instead of the reflectivities ($R_N$)
to prevent any artificial shifts in the reflectivities
(see Appendix~\ref{app:substrate} for more details).
The general feature of the spectral shapes 
can be elucidated as follows:
for thin samples with fewer than 40 layers, 
the contrast is subdued due to destructive interference from the substrate, resulting in 
a concave structure near $\lambda=520$ nm.
For thick samples with over 60 layers,
$R_N$ is predominantly influenced by contributions from the $N$-layer graphene and is minimally impacted by the substrate.
Consequently, given that $R_0$, suppressed by destructive interference, is in the denominator of the contrast, 
a convex structure appears near $\lambda=520$ nm.

The contrasts, calculated using Eq.~(\ref{eq:RN}),
are represented by the red solid curves in Fig.~\ref{fig:measured_fresnel}.
A satisfactory fit with no significant deviations is achieved
for various multilayer graphene samples, except for $C_1$ and $C_3$.
The observed discrepancy in these thin samples 
likely arises from differences in the interface between graphene and the substrate 
compared to that between adjacent graphene layers. 
This discrepancy is specific to thin samples and becomes negligible at 6 layers.
The notable agreement between measured and calculated reflectance for visible light 
clearly indicates the following two facts.
First, multilayer graphene can be treated as a collection of independent single layers,
which is plausible since 
the effect of interlayer stacking does not manifest in the dynamical conductivity
within the visible light range.
Second, reflectance measurements provide a reliable value of layer number $N$, averaged within a light spot. 
This is true even when the thickness of the graphene flakes determined by atomic force microscopy shows positional fluctuations
to a certain extent.
These facts will be used to validate the underlying assumption of our theoretical model
of reflection presented in the subsequent sections.

$C_{14}$ and $C_{15}$ vanish at $\lambda\simeq 2d_{\rm Si0_2}$, 
which has been referred to as zero reflection.~\cite{Anders1965,Roddaro2007a}
Zero reflection is useful in knowing the values of basic parameters later (at the end of Sec.~\ref{sec:sec3}).
To capture the essential role of graphene in achieving zero reflection, 
let us consider Eq.~(\ref{eq:RN}) when $\cos \varphi=0$.
For the numerator to vanish,
$i (n_{\rm Si}-n_{\rm SiO_2}^2) \cos \phi - 
\left( \frac{n_{\rm SiO_2}^2}{\sqrt{\varepsilon_{\rm g}}} - n_{\rm Si}\sqrt{\varepsilon_{\rm g}} \right) \sin \phi =0$
has to be satisfied.
Since $\phi$ is small, this equation can be simplified as 
$i (n_{\rm Si}-n_{\rm SiO_2}^2) -
\left( n_{\rm SiO_2}^2 - n_{\rm Si} \varepsilon_{\rm g} \right) \frac{d}{\lambda}2\pi N =0$,
which shows that $N\sim \frac{i d_{\rm SiO_2}}{\pi \varepsilon_{\rm g} d}(1-\frac{n_{\rm SiO_2}^2}{n_{\rm Si}})$ 
is an approximate layer number that gives zero reflection.
This argument makes it easy to understand that the dominant imaginary part of 
$\varepsilon_{\rm g}$ is essential for zero reflection to occur.

\subsection{Effective coupling constant}

As the difference between the calculated and experimental values of contrasts
proves to be sufficiently small for various multilayer graphene samples 
when utilizing the experimental $\varepsilon_g=(n+ik)^2$ values in Eq.~(\ref{eq:RN}),
the optical constants~\cite{El-Sayed2021}
are the results that theory should ultimately elucidate.
It can be inferred that nearly all the optical information of graphene multilayers 
is encompassed in the optical constant of monolayer graphene.
This inference is partially attributed to the fact that the effect of the stacking order 
on the reflectance does not manifest in the visible regime.
Consequently, we must inquire to what extent the theory of graphene optics can 
account for the $n$ and $k$ values.

Because of a conical energy-band structure of graphene known as the Dirac cone, 
the dynamical conductivity is well approximated by $\pi \alpha$ 
for visible light.~\cite{ando02-dc}
As a result, (suspended) monolayer graphene absorbs $\sim2.3$ percent ($=\pi \alpha$) 
of the incident visible light.~\cite{Nair2008}
A straightforward calculation of the Kubo formula shows that 
the dynamical conductivity of graphite is given by 
that of graphene divided by
the interlayer spacing $d$ ($=0.335$ nm):
$\sigma_{\rm graphite}=\pi \alpha/d$.~\cite{Kuzmenko2008,Sasaki2020a,Sasaki2020b}
The reflectance in the visible regime is free from 
the effects of the stacking order,~\cite{Taft1965,Ichikawa1966,Dresselhaus1976}
Fermi energy position, and temperature at room temperature ranges.~\cite{Klimchitskaya2018}
Thus, the relative permittivity of graphite for visible light wavelengths $\lambda$ 
is written as
\begin{align}
 \varepsilon_g = \varepsilon_r + i \frac{\alpha \lambda}{2d},
 \label{eq:varepsilon}
\end{align}
where $\varepsilon_r$ is the dielectric constant of the interlayer space.

\begin{figure}[htp]
 \begin{center}
  \includegraphics[scale=0.9]{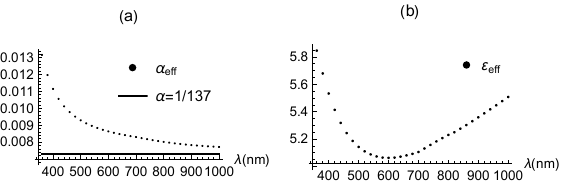}
 \end{center}
 \caption{{\bf Effective coupling constants.}
 (a) Plot of $\alpha_{\rm eff}$ calculated with the optical constants 
 obtained by El-Sayed {\it et al.}~\cite{El-Sayed2021}.
 The horizontal axis is $\lambda$ (nm).
 (b) Plot of $\varepsilon_{\rm eff}$.
 }
 \label{fig:epsilonG}
\end{figure}

By equating the right-hand side of Eq.~(\ref{eq:varepsilon}) with $(n+ik)^2$,
we define an effective coupling constant
$\alpha_{\rm eff} = 4nkd/\lambda$,
which is compared with $\alpha$ in Fig.~\ref{fig:epsilonG}(a).
The difference between $\alpha_{\rm eff}$ and $\alpha$ 
is actually larger than the order of 10 percent of $\alpha$ 
which is beyond the level of small corrections of order of $\alpha^2$ considered theoretically so far.
Similarly, we define $\varepsilon_{\rm eff} = n^2-k^2$ and plot it in Fig.~\ref{fig:epsilonG}(b).
If the interlayer space is a vacuum, an appropriate choice of $\varepsilon_r$ would be 1.
However, $\varepsilon_{\rm eff}$ is very different from unity
because the electronic wave function of the $\pi$-orbital spreads into the interlayer space,
light propagating in it is subjected to the spread of the wave function.~\cite{Fang2016,Rickhaus2020}
Fang {\it et al.}~\cite{Fang2016}
calculate $\varepsilon_r=6.9$ using a microscopic Poisson equation which has been tested by experiment.~\cite{Rickhaus2020}
The calculated effective dielectric thickness of graphene is found to be 0.22 nm,
and the microscopic dielectric permittivity decays from 6.9 in the carbon-atom plane 
to the vacuum permittivity within approximately 0.1 nm.
Though the calculation is for the electric field pointing in the direction perpendicular to the graphene sheet, 
similar (but slightly smaller) value is expected for the direction parallel to the sheet (let us assume it is $4 \sim 6$ here).

\section{Corrections to reflectance by light emission}\label{sec:sec3}

A substantial correction to the dynamical conductivity is deemed impermissible
because, if allowed, it would result in an inconsistency with the experimental 
observation that 
a (suspended) monolayer graphene absorbs approximately 2.3 percent ($=\pi \alpha$) 
of the incident visible light.~\cite{Nair2008}
An accurate theory capable of describing the reflectance does not necessitate
corrections to dynamical conductivity but rather requires 
a mechanism that explains reflectance without altering $\alpha$.
The model of light emission from graphene introduced in this section
serves as an illustration of such a mechanism.
Our model inherently incorporates the crucial concept of wave interference, 
specifically coherence or incoherence, 
as the initial phase of light emitted from each graphene layer.

\subsection{Basic idea}

\begin{figure}[htp]
 \begin{center}
  \includegraphics[scale=0.35]{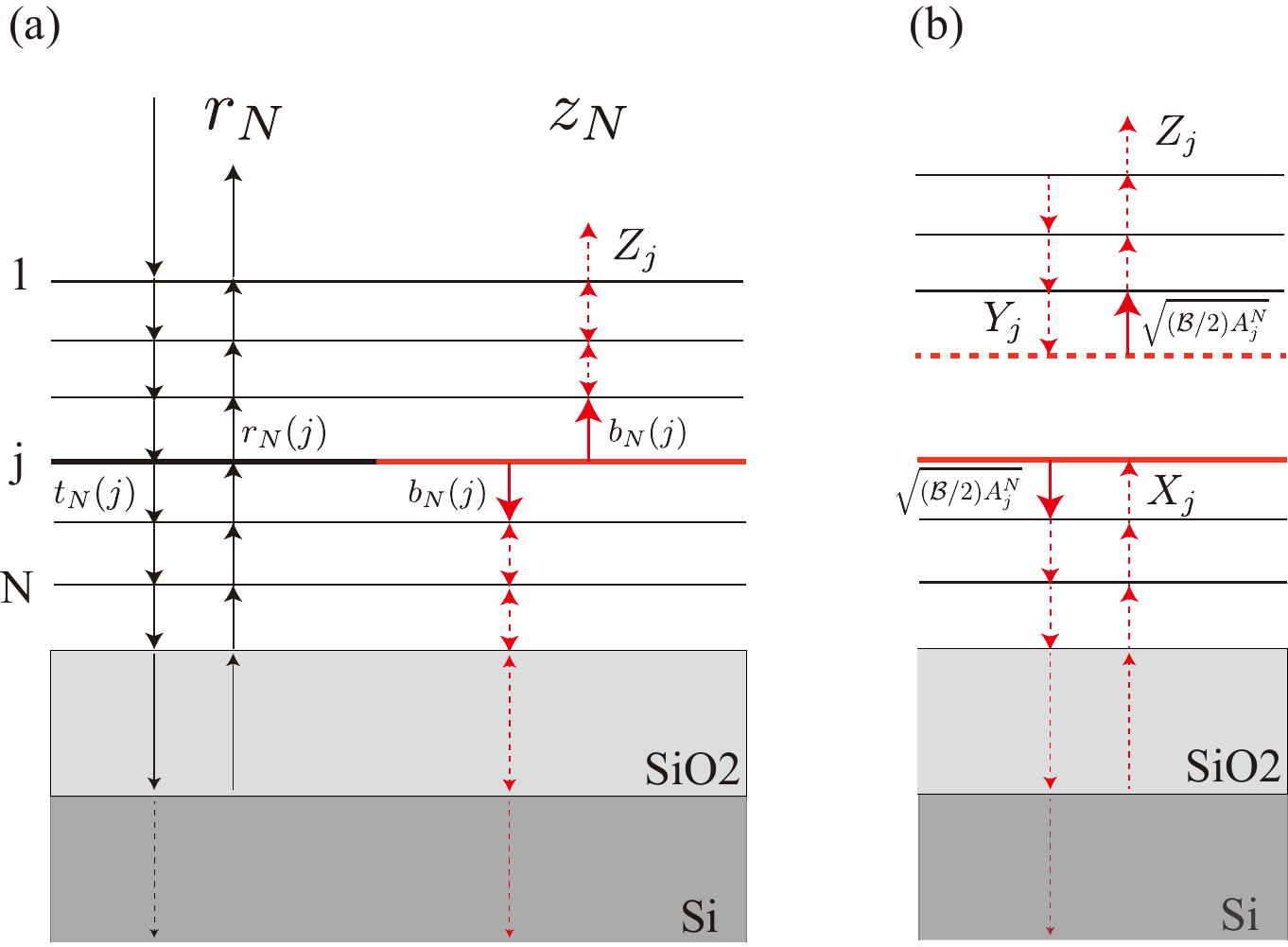}
 \end{center}
 \caption{{\bf Model description} 
 (a) The primary (left side) and secondary (right side) processes involved in the reflection 
 are physically interconnected through light absorption at each graphene layer.
 (b) The self-consistent calculation of $Z_j$ is explained in detail in the text.
 }
 \label{fig:LEmodel}
\end{figure}

Figure~\ref{fig:LEmodel}(a) illustrates our model of reflection,
where horizontal lines on the substrate represent $N$-layer graphene,
and the vertical lines depict light rays with arrows indicating the directions of light propagation.
The light rays on the left side (black in color) of Fig.~\ref{fig:LEmodel}(a)
show the primary processes of reflection (excluding contributions from light emission).
In this process, incident light from a light source is transmitted and reflected by graphene, 
while some energy of light being absorbed by each layer.
The reflection coefficient, $r_N$, is calculated from a primary model which is defined in Sec.~\ref{ssc:pm}.
The light rays on the right side (red in color) correspond to the light emission. 
Suppose that the $j$th layer emits light. 
The emitted light is transmitted and reflected by graphene until the light escapes the system,
and it contributes to the reflectance of the system. 
Thus, there is another ``reflection coefficient'' when $N$-layer graphene emits light 
which is defined in Sec.~\ref{ssc:model}.
Let $z_N$ denotes the sum over such amplitudes from all layers. 
Once we know what $r_N$ and $z_N$ are, 
then the reflectance is given by 
$R_N = |r_N + z_N|^2$.

We use a transfer matrix method to calculate $r_N$ and $z_N$.~\cite{Sasaki2020a,Sasaki2020b}
Transfer matrix method is useful in calculating 
reflection (up arrow) and transmission (down) coefficients at each layer
[$r_N(j)$ and $t_N(j)$ in Fig.~\ref{fig:LEmodel}(a)] in addition to 
the electric field $E_j^N$ ($j=1,\ldots,N$) that determines 
the absorption of the $j$th layer as
$A_j^N \equiv \pi \alpha |E_j^N|^2$.
The total absorption of $N$-layer graphene is 
$\sum_{j=1}^N A_j^N$.~\cite{Sasaki2020a}

\subsection{Primary model}\label{ssc:pm}

Basically, primary model means the Fresnel equation of Eq.~(\ref{eq:RN}) 
where $\varepsilon_g$ is given by Eq.~(\ref{eq:varepsilon}).
$\varepsilon_g$ has a single unknown constant $\varepsilon_r$ 
which can be estimated from the following observations.
Ultimately, we conclude that $\varepsilon_r=4.5$.

\begin{figure*}[htbp]
 \begin{center}
  \includegraphics[scale=1.]{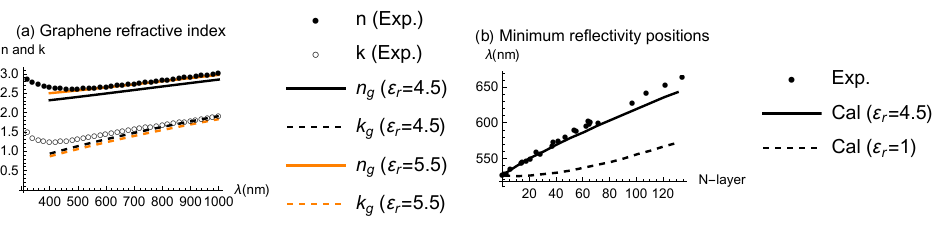}
 \end{center}
 \caption{{\bf $\varepsilon_r$ of the primary model}
 (a) Experimental $n$ and $k$ values are represented by dots, taken from Ref.~\onlinecite{El-Sayed2021}. 
 The lines depict bare optical constants without corrections. 
 (b) Dots indicate measured wavelengths corresponding to the minimum reflectance.
 A comparison between the measured and calculated results suggests that $\varepsilon_r=4.5$ is a reasonable value.
 }
 \label{fig:1}
\end{figure*}

First, there must not be a large discrepancy 
between $\varepsilon_g$ and the experimentally determined optical constants.~\cite{Bruna2009,El-Sayed2021}
Experimental values for $n$ and $k$~\cite{El-Sayed2021} are shown as $\bullet$ and $\circ$ in Fig.~\ref{fig:1}(a).
The lines depict (bare) optical constants plotted using Eq.~(\ref{eq:varepsilon}) with 
the refractive index $n_g={\rm Re}[\sqrt{\varepsilon_g}]$ and absorption coefficient
$k_g={\rm Im}[\sqrt{\varepsilon_g}]$ for $\varepsilon_r=4.5$ and 5.5.
When $\varepsilon_r=4.5$, $k\simeq k_g$ but $n$ has a certain difference from $n_g$.
When $\varepsilon_r=5.5$, $n\simeq n_g$ but $k$ has a certain difference from $k_g$.
Because $n_g$ increases with increasing $\varepsilon_r$ while $k_g$ decreases,
there is no $\varepsilon_r$ value that can reproduce $n$ and $k$ simultaneously.
This suggests that there should be such a discrepancy between them 
which is attributed to the corrections by light emission.
Second, $\varepsilon_g$ has to roughly reproduce the behavior of 
the reflectance of multilayer graphene. 
It exhibits a minimum at a certain wavelength, 
primarily due to destructive interference caused by SiO$_2$.
As shown in Fig.~\ref{fig:1}(b),
the position is red-shifted by increasing $N$,
indicating that even thin graphite samples significantly impact the light interference effect.
When $\varepsilon_r=1$, the position changes little, and 
a sizable artificial shift in wavelength is needed to ensure consistency
between theory and experiment, which cannot be explained as a correction.
When $\varepsilon_r=4.5$, a small difference between theory and experiment still remains.
However, as we show later,
the corrections provide better agreement 
not only for the wavelengths giving minimum reflectivity but also for the minimum reflectivity values,
thus accounting for the difference.
Third, $\varepsilon_r=4.5$ roughly reproduces the reflectivity of graphite 
in the infrared region.~\cite{Djurisic1999}
Similar $\varepsilon_r$ values have been used to 
reproduce the observed reflectivities of graphite and graphene.~\cite{Bruna2009,Castriota2019,El-Sayed2021}
We note that the value of $\varepsilon_r$ is less than the magnitude of the imaginary part of $\varepsilon_g$,
since visible light has a much longer $\lambda$ (400$\sim$800 nm) than $d$, although $\alpha$ is certainly a small quantity.
The optical properties of multilayer graphene are thus characterized by the large imaginary part of $\varepsilon_g$.

\subsection{Model of light emission}\label{ssc:model}


The corrections to the reflectance
are the main subjects of this paper.~\cite{Nemanich1980,Wang2008}
Specifically, we consider corrections where
some fraction of the energy absorbed by the $j$th layer
(of $N$-layer graphene) 
is transferred to light emitted from that layer [see the right side of Fig.~\ref{fig:LEmodel}(a)].
The amplitude of the emitted light is assumed to be the square root of the layer absorption
$A_j^N \equiv \pi \alpha |E_j^N|^2$,~\cite{Sasaki2020b}
multiplied by the branching ratio, ${\cal B}$, i.e., $\sqrt{\left( {\cal B}/2\right) A_j^N}$, 
where $1/\sqrt{2}$ means that the light emission is direction-independent
along the $c$-axis.
Note that $A_j^N$ depends not only on $j$ and $N$
but also on $\lambda$ and $d_{\rm SiO_2}$.

To examine how light emitted from the $j$th layer affects the reflectance, 
we define two subsystems, as shown in Fig.~\ref{fig:LEmodel}(b):
one is an isolated $(j-1)$-layer graphene in the air; 
the other is $(N-j)$-layer graphene on SiO$_2$/Si substrate.
Using the transfer matrix method, 
we can obtain the transmission and reflection coefficients 
of an isolated $(j-1)$-layer graphene in the air [denoted as $t_{j-1}^{g}$ and $r_{j-1}^{g}$]
and the reflection coefficient of $(N-j)$-layer graphene on the SiO$_2$/Si substrate
[denoted as $r_{N-j}$].~\cite{Sasaki2020a,Sasaki2020b}
Let the reflection coefficients be $X_j$ and $Y_j$ and transmission coefficient be $Z_j$
for the combined subsystems [see Fig.~\ref{fig:LEmodel}(b)].
These can be obtained by a self-consistent manner as follows.
After calculating $X^{(n)}_j$, we add it to $\sqrt{\left( {\cal B}/2\right) A_j^N}$ of the incident light to the 
$(j-1)$-layer graphene (in the air) as $\sqrt{\left( {\cal B}/2\right)A_j^N}+X^{(n)}_j$ and recalculate 
$Y^{(n+1)}_j=r_{j-1}^{g} \left( \sqrt{\left( {\cal B}/2\right)A_j^N}+X^{(n)}_j \right)$ 
and $Z^{(n+1)}_j=t_{j-1}^{g}\left( \sqrt{\left( {\cal B}/2\right)A_j^N}+X^{(n)}_j \right)$.
Then, we add a new $Y^{(n+1)}_j$ to $\sqrt{\left( {\cal B}/2\right)A_j^N}$ of the light incident to the 
$(N-j)$-layer graphene on the SiO$_2$/Si substrate as $\sqrt{\left( {\cal B}/2\right)A_j^N} +Y^{(n+1)}_j$ and recalculate 
$X^{(n+1)}_j=r_{N-j} \left( \sqrt{\left( {\cal B}/2\right)A_j^N}+Y^{(n+1)}_j \right)$.
These computations are repeated until $X_j$ and $Y_j$ converge.
In this way, we can obtain analytical expressions for the converged $X_j$, $Y_j$, and $Z_j$ for a given ${\cal B}$:
\begin{align}
 & X_j({\cal B})=r_{N-j} \left\{
 \frac{1+r_{j-1}^{g}}{1-r_{N-j} r_{j-1}^{g}} \right\}
 \sqrt{\left( \frac{\cal B}{2} \right) A_j^N}, \nn \\
 & Y_j({\cal B})=r_{j-1}^{g} \left\{
 \frac{1+r_{N-j}}{1-r_{N-j} r_{j-1}^{g}} \right\}
 \sqrt{\left( \frac{\cal B}{2} \right) A_j^N}, 
 \label{eq:hifo} \\
 & Z_j({\cal B})=t_{j-1}^{g} \left\{
 \frac{1+r_{N-j}}{1-r_{N-j} r_{j-1}^{g}} \right\}
 \sqrt{\left( \frac{\cal B}{2} \right) A_j^N}. \nn 
\end{align}

The ``corrected'' electric fields at an infinitesimal distance above and below the $j$th layer become
$\sqrt{\left( {\cal B}/2\right)A_j^N} +X^{(n)}_j + Y^{(n+1)}_j$ and $\sqrt{\left( {\cal B}/2\right)A_j^N} +Y^{(n)}_j + X^{(n+1)}_j$, 
respectively.
Self-consistency, whereby $\lim_{n\to \infty}X_j^{(n)}=X_j$ and $\lim_{n\to \infty}Y_j^{(n)}=Y_j$,
is therefore essential to ensuring that the corrected electric field
is continuous at the $j$th layer, which is a requirement of Maxwell equations. 
The corrected amplitude of the emitted light is written as
\begin{align}
 & \sqrt{\left( \frac{\cal B}{2} \right)A_j^N} +X_j + Y_j = \nn \\
 & (1+r_{j-1}^{g})
 \left\{ \frac{1+r_{N-j}}{1-r_{N-j} r_{j-1}^{g}} \right\}
 \sqrt{\left( \frac{\cal B}{2} \right) A_j^N}.
\end{align}
By comparing this with $Z_j({\cal B})$, we see that more accurate value of the amplitude of the emitted light 
is given by multiplying $\{ \cdots \} \sqrt{\left( {\cal B}/2\right)A_j^N}$ with $1+r_{j-1}^{g}$
as the renormalization constant, 
and $(1+r_{j-1}^{g})\{ \cdots \}\sqrt{\left( {\cal B}/2\right)A_j^N} $ is what $b_N(j)$ in Fig.~\ref{fig:LEmodel}(a) represents.
Therefore, we redefine $Z_j$ as 
\begin{align}
 & Z_j({\cal B}) \equiv t_{j-1}^{g} b_N(j), \label{eq:Znew} \\
 & b_N(j) \equiv (1+r_{j-1}^{g}) \left\{
 \frac{1+r_{N-j}}{1-r_{N-j} r_{j-1}^{g}} \right\}
 \sqrt{\left( \frac{\cal B}{2} \right) A_j^N}.
\end{align}

We can interpret $Z_j({\cal B})$ as follows.
The transmission coefficient $t_{j-1}^g$ is the direct propagation of the 
renormalized light emitted from $j$th layer to the air,
and $|t_{j-1}^g|^2$ monotonously decreases with increasing $j$.~\cite{Sasaki2020a}
The effects of scattering and absorption of the emitted light 
caused by surrounding layers is included by the part in the brace $\{ \cdots \}$.
It tends to suppress the magnitude of $Z_j$, but sometimes enhance.
For example, when $N=1$ (i.e., monolayer on a substrate), the part becomes $1+r_0$ 
which is larger than unity when $r_0$ is positive.

$Z_j({\cal B})$ is the value at zero initial phase, so the transmission coefficient can be given 
a phase degree of freedom expressing the coherence or incoherence 
of the light emission from the different layers:
\begin{align}
 z_N \equiv \sum_{j=1}^N e^{i\theta_j} Z_j({\cal B}).
 \label{eq:zn}
\end{align}
Accordingly, the corrected reflectance is uniquely determined by 
$R_N \equiv \left| r_N + z_N \right|^2 = \left|r_N \right|^2 + 2{\rm Re}\left[ r_N z^*_N \right] + \left|z_N \right|^2$.
The value of $z_N$ depends on these phases $\theta_j$.~\cite{Kuramoto1984}

We consider a case in which the phase is given by a coherent phase.
The exact derivation of the phase will be shown elsewhere
because it is beyond the scope of the present paper.
Here, we concisely explain the basic logic leading to the coherent phase
in terms of quantum electrodynamics.
First, 
we can define a quantum mechanical state of light 
($|\Psi_a \rangle$) that the primary model describes
(see left side of Fig.~\ref{fig:LEmodel}(a)).
All the information of light is expressed by the coefficients 
$r_N(j)$ and $t_N(j)$ ($j=1,\ldots,N$).
Second, we can also define another quantum state of light
($|\Psi_b \rangle$) for the emitted light
(see right side of Fig.~\ref{fig:LEmodel}(a)).
All the information of emitted light is expressed by the coefficients 
$b_N(j)$.
These two states have an overlap $b^*_N(j) t_N(j) + b^*_N(j) r_N(j)$
caused by $j$th layer graphene.
Thus, if we consider a linear superposition of these states as 
$|\Psi_a \rangle + e^{i\theta}|\Psi_b \rangle$ to form energy eigenstates,
the phase $e^{i\theta_j}$ must be chosen so that $e^{-i\theta_j} b^*_N(j) (t_N(j) + r_N(j))$ becomes a real number,
namely
\begin{align}
 e^{i\theta_j} = \pm \frac{t_N(j)+r_N(j)}{|t_N(j)+r_N(j)|} \frac{b_N(j)^*}{|b_N(j)|}.
 \label{eq:etheta}
\end{align}
The factor $\pm$ is a global phase ($\theta$)
in the sense that it is independent of the value of $j$.
Because the scattered light ($r_N$) and the emitted light ($z_N$)
form a two-level state, there are two possible
linear superpositions of their energy eigenstates,
$-1$ ($\theta=\pi$) or $+1$ ($\theta=0$).
The minus sign ($e^{i\pi}$) is assigned to the lower energy state.
From Eqs.~(\ref{eq:Znew}) and~(\ref{eq:etheta}),
we obtain $e^{i\theta_j} Z_j({\cal B}) = \pm \frac{t_N(j)+r_N(j)}{|t_N(j)+r_N(j)|} t_{j-1}^{g} |b_N(j)|$.

Including the correction due to coherent light emission leads to 
\begin{align}
 R_N = \left| r_N + \sum_{j=1}^N e^{i\theta_j} Z_j({\cal B}_{coh}) \right|^2,
\end{align}
where ${\cal B}_{coh}$ is the branching ratio of the energy of the emitted coherent photons 
to that of the absorbed photons.
Since coherent photon emission is related to the electron-photon coupling strength
of the annihilated photo-excited electron-hole pairs, 
${\cal B}_{coh}$ should be on the order of $(\pi \alpha)^2$ and insensitive to changes in $N$.

Next, we apply
$R_N = \left|r_N \right|^2 + 2{\rm Re}\left[ r_N z^*_N \right] + \left|z_N \right|^2$
to the case that $\theta_j$ in $z_N$  is a random variable.
Here, the definition of randomness is that if we take the time average regarding $\theta_j$, 
we have $\langle {\rm Re}\left[ r_N z^*_N \right] \rangle = 0$
and $\langle \left|z_N\right|^2 \rangle = \sum_{j=1}^N |Z_j({\cal B})|^2$.
We will refer to this case as incoherent corrections, which also include the cases that 
the global phase takes $0$ and $\pi$ if there is a perturbation that can mix the two energy levels.
An interference term is now included in $\left|z_N\right|^2$ as the last term of
\begin{align}
 \left|z_N\right|^2= \sum_{i=1}^N |Z_i({\cal B})|^2 + 
 \sum_{i\ne j} e^{i(\theta_i-\theta_j)} Z_i({\cal B}) Z^*_j({\cal B}),
\end{align}
but it vanishes when taking the time average
and only the first term of the incoherent corrections remains.~\cite{Velson2020}
Inelastic scattering of light such as 
Raman scattering is usually considered to give rise to incoherent photons.
Let ${\cal B}_{inc}$ be the branching ratio of the energy of the emitted incoherent photons to that of absorbed photons.
Since $Z_j({\cal B}_{inc})$ is proportional to $\sqrt{{\cal B}_{inc} A_j^N}$ [Eq.~(\ref{eq:hifo})], 
the incoherent corrections are proportional to ${\cal B}_{inc} A_j^N$.
For Raman scattering, the parameter ${\cal B}_{inc}$ is fundamentally determined by 
the electron-photon and electron-phonon coupling strengths,
and it should not be so sensitive to the change in $N$.
Indeed, the incoherent corrections with a constant ${\cal B}_{inc}$ follow 
the measured $N$ dependence of the $G$ band Raman intensity [Sec.~\ref{ss:raman}].
The $G$ band consists of optical phonons at the $\Gamma$ point, 
whose lattice vibrations are in-plane.

\begin{figure}[htbp]
 \begin{center}
  \includegraphics[scale=0.6]{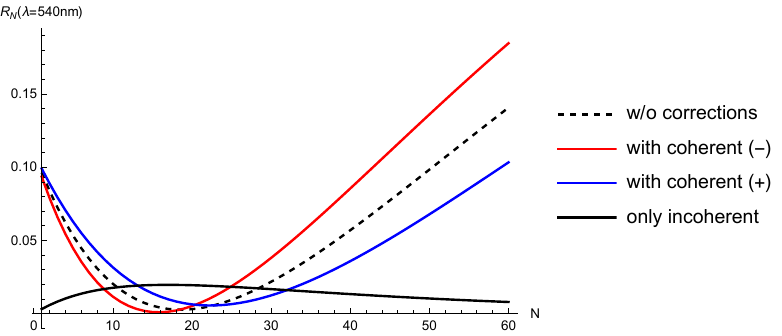}
 \end{center}
 \caption{{\bf Reflectance of $N$-layer graphene on SiO$_2$/Si substrate.}
 Dependence of the corrections to $R_N(540\ {\rm nm})$ [Eq.~(\ref{eq:formula})] on $N$, 
 where $d_{\rm SiO_2}=268$ nm, ${\cal B}_{inc}=0.1$, ${\cal B}_{coh}=0.0007$ and the global phase is $-$ or $+$.
 The incoherent components are enhanced (which is called interference-enhanced Raman scattering) at around 20 layers.
 }
 \label{fig:rs}
\end{figure}

A generalized reflection formula covering the above two cases
(coherent and incoherent corrections) can be written as
\begin{widetext}
\begin{align}
 R_N(\lambda,\theta,{\cal B}_{coh},{\cal B}_{inc}) 
 \equiv \left| r_N + e^{i\theta}\sum_{j=1}^N \frac{t_N(j)+r_N(j)}{|t_N(j)+r_N(j)|} \frac{b_N(j)^*}{|b_N(j)|} 
 Z_j({\cal B}_{coh}) \right|^2 
 +\sum_{j=1}^N |Z_j({\cal B}_{inc})|^2.
 \label{eq:formula}
\end{align}
\end{widetext}
When ${\cal B}_{coh}={\cal B}_{inc}=0$, $R_N(\lambda,\theta,{\cal B}_{coh},{\cal B}_{inc})$ reduces to Eq.~(\ref{eq:RN})
with Eq.~(\ref{eq:varepsilon}).
Figure~\ref{fig:rs} shows the $N$ dependence of $R_N(\lambda,\theta,0,0)$ (dashed),
$R_N(\lambda,\pi,{\cal B}_{coh}=0.0007,0)$ (red), $R_N(\lambda,0,{\cal B}_{coh}=0.0007,0)$ (blue), 
and $\sum_{j=1}^N |Z_j({\cal B}_{inc}=0.1)|^2$ (black),
for a fixed $\lambda=540$ nm.
Note that the incoherent corrections always increase the reflectance
and preclude zero reflections at $N\sim 15$, which is in contrast to the coherent corrections.
Moreover, the $G$ band Raman intensity is enhanced when zero reflection occurs.~\cite{Wang2008,Yoon2009a,Li2015,No2018} 
This situation is called interference-enhanced Raman scattering,~\cite{Nemanich1980}
and it is reasonably reproduced by Eq.~(\ref{eq:formula}).
The experimental fact that zero reflection is observed at $N\sim 15$ 
[see 14 and 15-layer in Fig.~\ref{fig:measured_fresnel}] shows that 
${\cal B}_{coh} \sim 0.0007$, 
${\cal B}_{inc}$ is much smaller than 0.1, and $\theta=\pi$ (red curve in Fig.~\ref{fig:rs}).

\section{Applications of model}\label{sec:sec4}

In this section, we show that
the discrepancy between 
the measured reflectance of multilayer graphene [Fig.~\ref{fig:measured_fresnel}]
and the prediction of the model [Eq.~(\ref{eq:formula})] is sufficiently small
for the present purpose.
Our model is, therefore, nearly equivalent 
to the Fresnel equation with the experimental optical constants ($n$ and $k$),~\cite{El-Sayed2021}
while our model can describe the interesting aspects of reflection.
Using monolayer graphene, we provide a detailed explanation of
the mechanism modifying the reflectance 
without introducing any artificial change in the dynamical conductivity.
To showcase the versatility of our model,
we also explore Raman scattering as incoherent corrections.

\begin{figure*}[htbp]
 \begin{center}
  \includegraphics[scale=1.0]{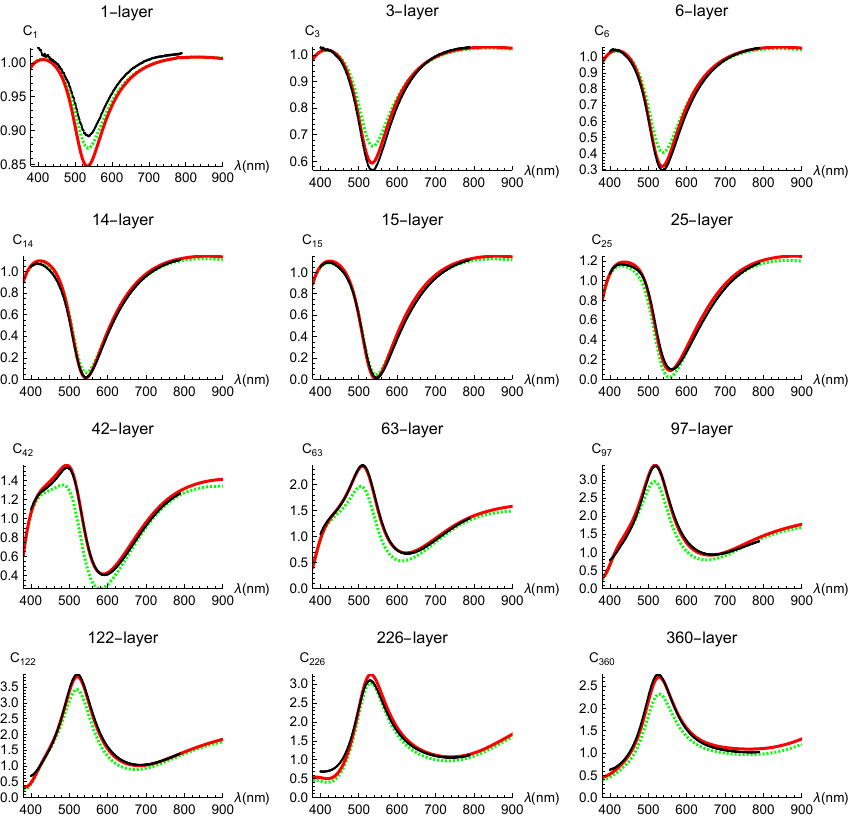}
 \end{center}
 \caption{{\bf Measured (black circles) and calculated (green dashed and red solid) spectral contrasts
 of $N$-layer graphene on the same SiO$_2$/Si substrate.}
 The green dashed curves represent the primary model [Eq.~(\ref{eq:RN}) with Eq.~(\ref{eq:varepsilon})]
 which does not include corrections. 
 The red solid curves include coherent corrections. 
 The horizontal axis is $\lambda$ (nm), and the reliable range of our spectrometer is 450 to 800 nm.
 }
 \label{fig:cn}
\end{figure*}

\subsection{Comparison of theory and experiment}

Calculated contrasts are shown in Fig.~\ref{fig:cn}
as red solid curves, which include coherent corrections only
(i.e., ${\cal B}_{inc}=0$).
Green dashed curves represent the primary model of Eq.~(\ref{eq:RN}), 
which does not include any corrections due to light emission.
All calculations were performed with 
${\cal B}_{coh}=0.0007$,
where this value was chosen so that we could obtain good agreement between the calculations and 
observations for all layers.
Note that this value is consistent with zero reflection being observed for $N\sim 15$.

From the consistency between the calculated and measured contrasts
shown in Fig.~\ref{fig:cn} (except $C_1$ and $C_3$),
we can draw two main conclusions.
First, the $\pi$ phase ($\theta=\pi$) of the coherent corrections is essential.
If we adopt 0 phase ($\theta=0$), a serious discrepancy arises, 
as can be readily imagined from the relative location of the red solid curves with respect to the green dashed ones. 
Second, the incoherent corrections are rather small.
In fact, for most of the layers examined (not shown in Fig.~\ref{fig:cn}), 
the incoherent corrections did not improve the fitting.
Our estimated reasonable range of ${\cal B}_{inc}$ is less than 0.01.

Only for the 226-layer,
there is a slight but non-negligible deviation of the red solid curve from the measured contrast.
A relatively small difference between the red solid and green dashed curves
shows that the strength of $|r_N+z_N|^2 - |r_N|^2$ is suppressed
and that $z_N$ is under some special phase balance by interference for $N\sim 226$.
Thus, a slight shift in $\theta_j$ might improve the fitting.
For example, second order corrections which arise due to a self-consistent calculation of $E_j^N$ (and $A_j^N$)
might be relevant to this.



\subsection{Monolayer}

Unfortunately, the Fresnel equation is inconsistent with 
the measured contrast of monolayer graphene (see $C_1$ in Fig.~\ref{fig:measured_fresnel}),
probably because reflectance depends on 
the condition of the interface between graphene and substrates.
However, as we have carefully confirmed that 
almost all the information of the corrections from the emitted light
is included in the reflectance of monolayer graphene,
we believe that the contrast $C_1$ 
calculated from the Fresnel equation with the experimental $n$ and $k$ values
is the result that we should compare with the model.

\begin{figure*}[htbp]
 \begin{center}
  \includegraphics[scale=1.0]{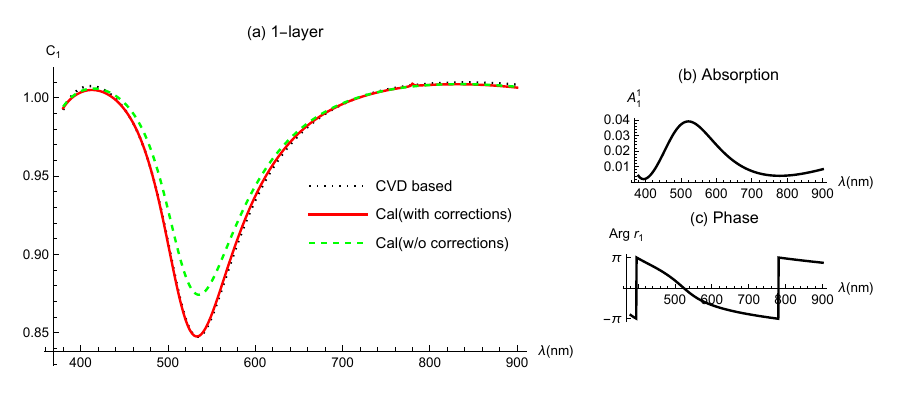}
 \end{center}
 \caption{{\bf 
 Contrasts of monolayer graphene on SiO$_2$/Si substrates.}
 (a) One contrast is calculated using Eq.~(\ref{eq:RN}) with the experimental $\varepsilon_g$ (black dotted),
 while the other two curves are obtained using the model with corrections (red solid curve)
 and without corrections (green dashed curve).
 The increased visibility of graphene is primarily attributed to substrate-induced 
 enhancement of light absorption and partly to light emission.
 (b) Layer absorption of monolayer graphene on SiO$_2$/Si substrates $A_1^1 = \pi \alpha |1+r_1|^2$.
 The destructive interference effect of the substrate enhances absorption near $\lambda\simeq 2d_{\rm SiO_2}$,
 where $r_1$ is a positive number (c).
 }
 \label{fig:11}
\end{figure*}

In Fig.~\ref{fig:11}(a),
we present simulated (black dotted) and calculated (green dashed and red solid) spectral contrasts
of monolayer graphene on the SiO$_2$/Si substrate.
Clearly, the corrections are of physical significance;
black dots and red solid curve almost perfectly match.
In $R_1=|r_1 + z_1|^2$, 
$r_1$ becomes a positive number only near $\lambda=2d_{\rm SiO_2}$
as shown in Fig.~\ref{fig:11}(c).
This is due to the destructive interference caused by SiO$_2$,
which also increases the absorption because $A_1^1 = \pi \alpha |1+r_1|^2$ [Fig.~\ref{fig:11}(b)].
This enhanced absorption leads to the main difference between the reflectances from the substrate ($R_0$) and 
from the graphene on it ($R_1$), increasing the visibility of graphene.
The correction due to light emission $z_1$ 
is a negative number due to the negative global phase of an energetically stable configuration of light.
Thus, the corrections increase $|R_1-R_0|$.
Namely, the increase in the visibility of graphene is mainly due to the substrate-induced 
enhancement of light absorption and is partly due to light emission.

It is important to note that 
the primary model can explain the measured contrasts if $\alpha$ is more than 20 percent larger
than $1/137$.
However, this immediately leads to an inconsistency with the experimental fact that 
(suspended) monolayer graphene absorbs $\sim2.3$ percent ($=\pi \alpha$) 
of the incident visible light.~\cite{Nair2008}
Additionally, the primary model with such corrections to $\alpha$ does not reproduce 
the measured contrasts of many samples with different thickness.
Furthermore, the primary model is not applicable to the Raman effect, 
while our model can include it in a natural way, as shown below.

\subsection{Raman scattering as incoherent corrections}\label{ss:raman}

\begin{figure}[htbp]
 \begin{center}
  \includegraphics[scale=0.7]{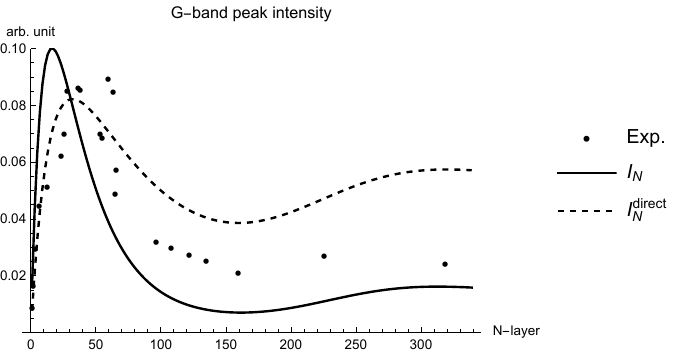}
 \end{center}
 \caption{{\bf Raman intensity as incoherent corrections.}
 The dots represent the measured peak intensity of the Raman $G$ band when using a light source with a wavelength of 532 nm.
 The dashed curve corresponds to the incoherent correction, 
 while the dot-dashed curve is an approximation that ignores multiple scattering.
 }
 \label{fig:2}
\end{figure}

We measured the $G$ band Raman peak intensity as a function of layer number
in order to verify the interpretation of Raman scattering being an incoherent light emission.~\cite{Wang2008}
The incoherent correction is defined by the last term of Eq.~(\ref{eq:formula}) as
$I_N \equiv \sum_{j=1}^N |Z_j({\cal B}_{inc})|^2$.
$I_N$ is proportional to ${\cal B}_{inc}$
and does not depend on the value of ${\cal B}_{inc}$ when scaled.
As shown in Fig.~\ref{fig:2}, 
there is a reasonable similarity between the measured Raman intensity (dots) and calculated incoherent component
(dashed curve).
Also plotted is an approximation of $I_N$ (dot-dashed curve) defined by
\begin{align}
 I^{\rm direct}_N \equiv \frac{{\cal B}_{inc}}{2} 
 \sum_{j=1}^N |t_{j-1}^{g}|^2 A_j^N,
\end{align}
to show the effect of multiple scattering of incoherent light. 
The approximation overestimates the intensity for thick samples, as readily imagined.

On the other hand, 
there is a noticeable discrepancy between them for samples with fewer than 30 layers,
where there is a dip in the reflectance that is similar to the observation by No {\it et al}.~\cite{No2018}
The assumption of a random phase for $\theta_j$ in Eq.~(\ref{eq:zn})
is a possible reason for the discrepancy, 
because random phases can undergo synchronization or entrainment.~\cite{Kuramoto1984}
An intermediate state of the phase $\theta_j$ that is neither random nor perfectly coherent
may account for the behavior.

\section{Discussion}\label{sec:dis}

There is a possibility that substrates play a decisive role in determining
the selection of the two states specified by the global phase $\theta$ (0 and $\pi$).
To see this, let us consider monolayer graphene suspended in the air.
From Eq.~(\ref{eq:RN}), 
the reflection and transmission coefficients (without substrates) are 
$r_1 \sim - \pi\alpha/2$ and $t_1 \sim 1-\pi \alpha/2$,
and the absorption is given by 
$A_1^1 =1-r_1^2 -t_1^2 \sim \pi\alpha$.
Light emission modifies the reflection coefficient as follows:
\begin{align}
 r_1 + e^{i\theta} Z_1({\cal B}) = 
 - \frac{\pi \alpha}{2} + 
 e^{i\theta} \sqrt{\frac{{\cal B}}{2} \pi\alpha}.
\end{align}
The magnitude of the second term is 0.0028
when ${\cal B}=0.0007$ which is about 25 percent the magnitude of the first term (0.011).
The reflection increases or decreases depending on $\theta$.
When $\theta=\pi$ (0), the correction term is negative (positive) in sign;
therefore, the light emission increases (decreases) the reflectance.
Mathematically speaking, 
the change in the reflectance is equivalent to a replacement of $\alpha$ as
$\alpha \to \alpha_{\rm eff} = \alpha -e^{i\theta} \sqrt{\pi{\cal B}/2\alpha}$.
This immediately leads to an inconsistency with the experimental fact that 
suspended monolayer graphene absorbs $\sim2.3$ percent ($=\pi \alpha$, 
the measured uncertainty is within 5 percent of $\pi \alpha$) 
of the incident visible light.~\cite{Nair2008}
This inconsistency is resolved by considering that for graphene suspended in the air
the states of $\theta=\pi$ and 0 are degenerate and the effect of light emission effectively 
disappear by interference.
This contrasts with monolayer graphene on SiO$_2$/Si substrates
for which $\theta=\pi$ is selected.


Since coherent and incoherent emissions are 
two extreme cases (uniform and random) of the phase $\theta_j$ in Eq.~(\ref{eq:zn}),
a sharp distinction between the coherent and incoherent emissions is not always possible.
The proper way to calculate the reflectance is to derive 
a dynamical model of $\theta_j$ at a microscopic level,~\cite{Dicke1954,Kuramoto1984}
and use it to calculate
$R_N = \left| r_N + \sum_{j=1}^N e^{i\theta_j} Z_j({\cal B}) \right|^2$.
Especially in the case of monolayer, they are inseparable as
\begin{align}
 R_1 = \left| r_1 + e^{i\theta_1} Z_1({\cal B}_{coh}) + e^{i\phi_1} Z_1({\cal B}_{inc})\right|^2,
\end{align}
when $\phi_1$ and $\theta_1$ have some correlation.
Then, the reflectance is always subject to fluctuations created by the last term (through electron-phonon interactions).

Our model postulates that 
the effect of the emitted light appears as a correction to the reflection (and transmission) coefficient, 
not as a correction to the dynamical conductivity.
Meanwhile, spontaneous emission is generally treated as a loss, 
and it is often included as a phenomenological relaxation constant in the dynamical conductivity. 
For example, in Ref~\onlinecite{El-Sayed2021}, the authors introduce relaxation constants 
for Drude-Lorentz oscillator model to interpret the measured optical constants. 
However, an excessively large relaxation constant (or very short lifetime) for the Drude term (0.6 fs) 
already raises concerns about the naturalness of including such a relaxation parameter.~\cite{Toqeer2021}
The justification of our postulate needs a theoretical clarification 
at a more fundamental level of quantum electrodynamics, 
which is capable of describing photon creation and annihilation,
while the excellent agreement between the model and experiments clearly shows that 
this postulate works well. 
Our model is also consistent with a theoretical result that 
the dynamical conductivity is free from such a correction 
when graphene is undoped (i.e., charge neutrality condition is satisfied).~\cite{ando02-dc}
Moreover, the model explains the $N$-dependence of the Raman intensity 
as the incoherent corrections (Sec~\ref{ss:raman}), 
besides the contrast, in a unified manner.

\section{Conclusion}\label{sec:con}

In summary, we succeeded in explaining the measured visible contrasts of multilayer graphene samples on 
an SiO$_2$/Si substrate by including coherent light emissions that come from some fraction 
(${\cal B}_{coh}=0.0007$) of the absorbed photon energy.
The coherent corrections are essential for assuring the consistency between theory and experiment, 
while the incoherent corrections can be neglected for the contrast.

Photo-excited electrons contribute insignificantly to light emission 
when they are distant from the bottom of the conduction band.
Namely, the value of ${\cal B}$ for those electrons would be suppressed.
In fact, graphene lacks a bandgap and 
the branching ratio of coherent light emission to absorbed light is very small (${\cal B}_{coh}=0.0007$). 
What we have argued for in this paper is corrections (to the main effect) that have small branching ratios.
However, whether light emission from those electrons can be entirely neglected 
depends on various factors.
Graphene serves as an interesting example where 
corrections are greatly enhanced by its large absorption.
The destructive interference effect from the substrate and the multilayer-induced coherence are 
the means by which it is observable in the reflectance.
Similar emission-based corrections could be anticipated for other layered materials without band gaps, 
and the method developed here may prove useful in accurately understanding their optical properties. 

Our formulation of the reflectance 
using the transfer matrix method has a descriptive ability for layered materials
having defects and irregularities.
For this reason, and considering the success it has had in describing 
the reflectance of relatively simple systems (graphene multilayer at visible range), 
we believe that some future form of this theory may be useful 
in describing the optical properties of any layered material with or without a band gap.

\section*{Acknowledgments}

K. S thanks Y. Sekine, H. Endo, and Y. Taniyasu for raising helpful questions on this subject.
Part of this work was supported by ``Nanotechnology Platform Japan'' of the Ministry of Education, 
Culture, Sports, Science and Technology (MEXT), Grant Number JPMXP09F21UT0045.
The reflectance measurement was conducted in the Takeda Cleanroom with help of the
Nanofabrication Platform Center of the School of Engineering, the University of Tokyo, Japan.

\

\appendix

\section{Measured reflectivities}\label{app:substrate}

\begin{figure*}[htbp]
 \begin{center}
  \includegraphics[scale=1.0]{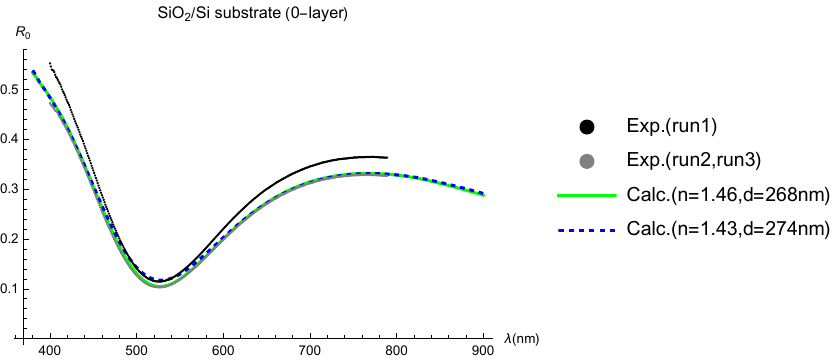}
 \end{center}
 \caption{{\bf Determination of SiO$_2$ thickness.}
 Measured (black [run 1] and gray [run 2 (run 3)] dots) and calculated (green and blue dashed curves) 
 spectral reflectivities of $N$-layer graphene on the same SiO$_2$/Si substrate.
 A white-light source was used, and the reflectivity was measured at room temperature. 
 The horizontal axis is $\lambda$ (nm), and the reliable range of our spectrometer is 450 to 800 nm.
 }
 \label{fig:r0}
\end{figure*}

We performed the reflectivity measurements three times on the same sample (run 1, run 2, and run 3).
All measurements were consistent in the sense that the measured reflectivities of $N$-layer graphene 
on the substrate ($R_N$) divided by that of the SiO$_2$/Si substrate ($R_0$), 
that is, the contrast ($C_N = R_N/R_0$), were identical.
Therefore, we compared $C_N$ with calculations in the main text.
However, while $R_0$ of run 2 and run 3 were identical, 
$R_0$ of run 1 deviated from that of run 2 (run 3), as shown in Fig.~\ref{fig:r0}.
The same contrast also means that the values of $R_N$ for run 2 and run 3 were identical, 
while $R_N$ for run 1 deviated from that of run 2 (run 3).
The discrepancy between $R_0$ of run 1 and that of run 2 (run 3)
was presumably brought about by a change in the focal point along the depth direction
when a reference substrate was replaced with the target sample.
This could change the incident light intensity 
and lead to a discrepancy in $R_N$ between run 1 and run 2 (run 3).

A problem arises when we determine the value of $d_{\rm Si0_2}$ from $R_0$.
Considering that the wavelength giving the reflectivity minimum is the same for the three runs, 
we can determine $d_{\rm Si0_2}=268$ nm by assuming that $n_{\rm Si0_2}$ is the standard value ($\simeq$1.46).~\cite{Malitson1965}
These parameters reasonably reproduce $R_0$ of run 2 (run 3) [Fig.~\ref{fig:r0}].
Meanwhile, 268 nm is inconsistent with another estimation using a reflectometer ($274 \pm 1$ nm), which results in 
$n_{\rm Si0_2}\simeq 1.43$ to reproduce the wavelength giving the reflectivity minimum by destructive interference
(the same $n_{\rm Si0_2} d_{\rm Si0_2}$). 
When we choose $n_{\rm Si0_2}\simeq 1.43$ for $d_{\rm Si0_2}=274$ nm, 
the corresponding $R_0$ is the (blue) dashed curve, which reproduces $R_0$ of run 1 around the reflectivity minimum 
(the two curves overlap from 460 to 560 nm)
and approaches $R_0$ of run 2 (run 3) away from the reflectivity minimum
($\lambda < 460$ nm and $\lambda > 560$ nm).
Since $n_{\rm Si0_2}\simeq 1.43$ is an acceptable value, 
we need to mindful of the possibility that 
a true $R_0$ is neither $R_0$ of run 1 nor that of run 2 (run 3).
Because we have assumed that the refractive index of Si is the commonly used value, 
we keep the same standpoint for SiO$_2$.
Ultimately, we concluded that $d_{\rm Si0_2}=268$ nm.



\bibliographystyle{apsrev4-2}

%

\end{document}